\documentclass{aastex}
\usepackage{spr-astr-addons}
\usepackage{url}

\RequirePackage{color}

\begin{document}

\title{The search for Primordial Quark Nuggets among Near Earth Asteroids}

\shorttitle{Quark nuggets among NEAs} \shortauthors{Horvath}

\author{J. E. Horvath}
\affil{Instituto de Astronomia, Geof{\'\i}sica e Ci\^encias
Atmosf\'ericas, Universidade de S\~ao Paulo, Rua do Mat\~ao 1226 ,
05508-090 S\~ao Paulo, SP, Brazil}

\begin{abstract}
Primordial Quark Nuggets,remnants of the quark-hadron phase
transition, may be hiding most of the baryon number in superdense
chunks have been discussed for years always from the theoretical
point of view. While they seemed originally fragile at
intermediate cosmological temperatures, it became increasingly
clear that they may survive due to a variety of effects affecting
their evaporation (surface and volume) rates. A search of these
objects have never been attempted to elucidate their existence. We
discuss in this note how to search directly for cosmological
fossil nuggets among the small asteroids approaching the Earth.
``Asteroids'' with a high visible-to-infrared flux ratio, constant
lightcurves and devoid of spectral features are signals of an
actual possible nugget nature. A viable search of very definite
primordial quark nugget features can be conducted as a spinoff of
the ongoing/forthcoming NEAs observation programmes.

\end{abstract}



\section{Introduction}

Primordial quark nuggets (PQNs) have been postulated to contain
most of the baryonic number of the universe many years ago
(Witten, 1984), being a physical realization of the so-called {\it
strange matter} hypothesis. The strange matter hypothesis states
that a cold, catalyzed form of the quark gluon plasma could be the
true ground state of the hadronic matter and, if formed at the QCD
scale when color become confined, remnants of substellar mass
could help to "hide" the baryon number content and achieve a large
value of $\Omega_{matter} \sim 0.25$ now favoured by observations.
The initial excitement about this possibility gave place to
several analysis addressing whether the PQN could survive from the
initial high-temperature environment until the lower temperature
universe, where the contribution $- T S$ to their free energy make
them stable against evaporation-boiling. In fact both processes
(evaporation of ordinary hadrons from the surface and bulk
conversion -boiling-) have been examined with varying results
(Madsen and Olesen, 1991; Alcock and Olinto, 1989; Olesen and
Madsen, 1993; Sumiyoshi and Kajino, 1991). Generally speaking, it
can be stated that the slower the process, the larger the mass
that can survive; thus it is of great interest to pinpoint the
realistic models to estimate reliably a surviving mass. It is
generally agreed that the latter can not be larger than the mass
inside the horizon, corresponding to a baryon number of $A_{hor} =
10^{49} ({100 \, MeV\over{T_{qcd}}}) $ (where $T_{qcd}$ is the
temperature for the confinement phase transition, assumed to be
first order); and while no consensus has been achieved on the
exact value of this surviving nugget mass, values as low as $A =
10^{40}$ have been estimated in the literature (Sumiyoshi and
Kajino, 1991; Bhattacharyya et al., 2000). Recently we have shown
that the inclusion of pairing effects in strange matter would
increase the binding energy of the nugget and hence push down the
survival mass substantially (Lugones and Horvath, 2004). There is
a strong belief that dense quark matter should undergo pairing of
some of all quark flavors, although the precise form of the phase
diagram is still under debate (Alford, 2006).

Given this attractive possibility we address here a novel and
feasible search for PQN in the asteroidal-mass range; in several
senses complementary to the searches of smaller masses
(nuclearities) already performed or underway (Finch, 2006). We
present a discussion of the general features of the nuggets in
Section 2. Section 3 addresses the detectability of this
population by photometric and spectroscopic techniques. A
discussion and some conclusions are summarized in last Section.

\subsection{Primordial Quark Nuggets}

According to the Bodmer-Witten-Terazawa hypothesis (Witten, 1984;
see previous work by Bodmer, 1971 and Terazawa, 1979), a stable
form of the QGP with high strangeness content could be self-bound
and therefore form objects from a minimum (nuclear) size until
$\sim \, M_{\odot}$ at the onset of the general relativistic
instability. While the former (``strangelets") are amenable to
particle search techniques (Klingenberg, 1999) and the latter may
constitute compact stars (Witten, 1984; Alcock, Farhi and Olinto,
1986; Haensel, Zdunik and Schaeffer 1986; Benvenuto and Horvath
1989), in the intermediate range PQNs would resemble extremely
compact asteroids. For constant density $\rho = 5 \times 10^{14} g
\, cm^{-3}$ (which is an excellent approximation for this range;
see Alcock, Farhi and Olinto, 1986) the size of the nuggets is

\begin{equation}
R_{N} = 2 \times 10^{2} \, \times \, {\biggl[
{\frac{M_{N}}{10^{-12} M_{\odot}}} \biggr]}^{\frac{1}{3}} \, cm
\end{equation}

and we have scaled to the mass corresponding to a baryon number $A
= 10^{45}$ above the evaporating mass (Lugones and Horvath, 2004).

Because of the large size of the galactic halo, the isotropic
nugget flux of this mass-scale is always very small, of the order
of $3 \, \times \, 10^{-37} cm^{-2} s^{-1} sr$ onto the Earth
neighborhood. For a minimum approach of one PQN at a distance $D
\leq \, 2 \times$ the Earth-Moon distance, or $2 \times R_{Moon}$
(see below) a passage would happen at a rate $10^{-6} \, yr^{-1}$.
This small rate gives no hope for detecting PQNs freely roaming
the halo passing by in open orbits.

A {\it captured} nugget population in orbit provides much better
prospects. To be captured by the gravitational field of the Sun
the nuggets coming from an isotropic flux must loose energy and
angular momentum. A first viable mechanism would be a number of
flyby encounters with stars in the galactic disk. Then a fraction
of these slower nuggets, having now $E \gtrsim 0$ respect to the
solar potential would be captured in fly-by encounters with
Jupiter. Let us stress that even if the distribution of PQNs will
not slow down {\it statistically}, {\it individual} objects from
that distribution may be effectively captured by this mechanism.
The PQNs with this mass-scale do not slow down appreciably even by
passing through stars, since the basic energy loss rate
$dE/dt=-\pi R_{N}^{2} \rho_{matter} v_{N}^{3}$ (with
$\rho_{matter}$ the density of the traversed stellar material and
$v_{N}$ the nugget velocity $\sim \, 250 \, km \, s^{-1}$) is many
orders of magnitude smaller than the incoming kinetic energy of
the nugget.

The estimated number of PQNs effectively captured is always tiny,
but perhaps enough to yield 10-100 of them bound to the solar
system out of the $\sim 10^{23}$ existing in the whole galactic
population. Note that this capture rate, if extrapolated to all
the stars in the disk, would still mean a total capture nugget
fraction of $\leq \, 10^{-10}$. A variety of orbital elements is
expected because of the random incidence, however, since the
nuggets would have Apollo-like, planet-crossing orbits, their
residence in the solar system would be limited to a dynamical time
of $\leq 10 \, Myr$. Finally, there should be a huge number ($\geq
10^{7}$) of PQN at a given time inside the bounds of the solar
system, adding up to a mass comparable to several times the
Earth's mass. They may be perturbed and directed to the inner
solar system much in the same way that ordinary cometary nuclei
do, only that they would remain unseen most of the time. The
identification of the tiny dense nuggets among the subpopulation
of NEAs would be the subject of the next section.

\section{How to detect nuggets among NEAs}

A great deal of attention has been paid recently to the issue of
NEAs and the possible hazards for the Earth in the case of direct
collisions. Actually, a few observing programs to look for
close-by encounters are already operating (for example,
Spacewatch, see Mc Millan, 2006) or being implanted. A huge number
of small NEAs, those not very hazardous to the Earth in the
ballpark of $\sim \, tens$ of meters are being currently detected.
The observations indicate that at a given time $> 100$ asteroids
pass closer than the Moon, while at least one of $\sim 30 \, m$
and $\simeq 10$ of about $10 \, m$ collide with the Earth each
year. At a distance comparable to the Moon these would have
magnitudes $m_{v} = 13$ and $15.5$ respectively (assuming, as
usual, a zero phase angle; see Paczy\'nski, 2006). These
fast-moving boulders would be visible for a $\sim \, days$ at most
because of their high velocities relative to the Earth, of the
order of $20 \, km \, s^{-1}$.

As is well-known, one of the simplest forms for determining the
albedo $A$ (and hence gather information on their composition) of
an asteroid is to perform at least simultaneous $V$ and $I$ images
to infer $A$ from the quotient of fluxes. The exotic nature of the
nuggets allows one relatively easy form of distinguishing them
from conventional asteroids: since the strange quark matter is
expected to have a plasma frequency as high as $20 \, MeV$ (well
in the hard $\gamma$-ray frequencies), the bare quark surface
would act as a perfect mirror to the incident solar light. Hence,
contrary to the case of even metallic asteroids for which $A \sim
0.1$, we expect albedos $\approx \, 1$ and therefore a quotient
$F_{V}/F_{I}$ much larger than any reasonable normal surface.
Therefore a PQN would appear to have a magnitude $m_{V} \simeq \,
20$ at a distance $\sim \, R_{Moon}$, looking like a larger
``normal'' asteroid but showing the abnormal flux quotient.
Specific programmes like the Spacewatch (Mc Millan, 2006) already
mentioned and several other programs (see for example the
compilation of the NASA, 2007) currently monitor objects fainter
than $m_{V} \sim \, 20$, and should be able to detect $\sim \, 1
\, m$ nuggets out to $\sim \, R_{Moon}$ (hence the distance
estimate given in the former Section). A recent discussion by
Paczy\'nski (2006) has shown the convenience of a satellite at the
L1 Lagrangian point for the NEAs study and early alert. The main
difficulties for such identifications are the high proper motion
of the candidates, which render them observable for a short time
$\sim 1 \, day$, and also the large number of candidates in a
given night, provided that there is no simple way of selecting a
subsample, for instance, based on orbits.

A second prediction for these objects is that their spectra would
be essentially solar and devoid of any characteristic line that
distinguish asteroidal compositions. This seems promising,
although spectroscopy of dim fast-moving objects is certainly
difficult. In any case the slower candidates (say, with $\mu \leq
1"/s$) could be scanned by suitably setup devices. Thus both the
photometric and spectroscopic measurements would show signatures
of the exotic nugget features difficult to miss for good
observational data. This feature (and the former flux quotient as
also) implicitly assumes that the surface of the nugget is {\it
not} covered with normal matter, which would restore the ability
of radiating conventionally the reflecting light. The situation is
quite like the discussion of strange stars (Alcock, Farhi and
Olinto, 1986; Haensel, Zdunik and Schaeffer 1986; Benvenuto and
Horvath 1989), but now referring to much smaller objects, 12
orders of magnitude less massive than a strange star, with a
correspondingly small surface gravity. This consideration
supports, but does not prove, the expectation of a bare surface.

A third complementary signature could be obtained by precise
stellar occultation observations, given that the tiny nugget with
high albedo would mimic a larger radius asteroid (at least 10
times bigger) but would show a very short occultation time
corresponding to a small and sharp edge ball. Even the {\it lack}
of occultation where it should be one (if the asteroid was normal)
could be helpful for characterizing the PQN candidates.  Finally
we note that a very uniform light curve, (in contrast with most of
the asteroids of irregular size which tumble and rotate) is
expected from the spherical superdense nugget, quite independently
of the wavelength. This completes a short list of testable
predictions for these objects.

Several contemporary sources (as opposite to the discussed
primordial population) that could supply a flux of small nuggets
have been advanced, like the disruption of binary strange stars
(Madsen, 2005; however see Klu\'zniak and Lee, 2002 for a
calculation of a strange star-black hole encounter not leaving any
ejected debris). It is amusing to note that long ago Friedman and
Caldwell (1991) predicted the stripping of nuggets of precisely $A
= 10^{45}$ from the simple balance of tidal and surface forces in
a binary merging. However, the small number of events of this type
$\sim 10^{-5} yr^{-1}$ argues against stripped nuggets of this
origin being a significant source when compared to the
hypothetical primordial population, and in any case there is no
way for distinguishing both.

The possibility of a {\it direct impact} onto the Earth is
extremely small (about one event per Hubble time) for halo PQNs,
but grows considerably for a captured population. Specific
signatures of such an hypothetic collision (likely giving rise to
a huge epilinear earthquake) have never been worked out in full
detail. However, lighter PQNs from astrophysical origin may be
candidates to direct impacts with masses down to tens of tons (see
Herrin, Rosenbaum and Teplitz, 2006, for this milder but related
signal search).

\section{Conclusions}

We have discussed some observations which may possibly test the
presence of a population of primordial quark nuggets captured in
solar orbits. We have been led by the fact that the mass range
predicted by theoretical calculations falls precisely in the
ballpark of small asteroids, and by the existence of advanced
programmes of NEAs search. We suggested a few observational tests
to reveal the nature of the nuggets as they approach the Earth. If
an efficient setup for the fast-moving candidates can be
established it would be possible to identify some definite
signatures predicted by the basic physics of the nugget surface
reflecting sunlight. Although the search for PQNs could be
difficult and take several years, the payoff would be uttermost
rewarding even if they are not finally found. In any case,
valuable information about the normal asteroid population would be
gathered. Few other programmes could be suitable for searching
such an elusive primordial relics.

\acknowledgments CNPq (Brazil) is acknowledged for financing JEH
through a fellowship. C. Beaug\'e, F. Roig and S. Ferraz-Mello
have provided scientific advice to this work. Finally, the
criticisms of an anonymous referee have added a few interesting
points to the present article.

\end{document}